\documentclass{svjour3}                     
\usepackage{graphicx}  
\usepackage{mathptmx}
\usepackage{amsmath}
\usepackage{bm}

\journalname{Journal of Statistical Physics}

\begin{document}

\title{Partially Annealed Disorder and Collapse of Like-Charged Macroions}

\author{Yevgeni Sh. Mamasakhlisov \and Ali Naji \and Rudolf Podgornik}
\authorrunning{J. Mamasakhlisov \etal}

\institute{
 Yevgeni Sh. Mamasakhlisov 
 \at 
 Dept. of Molecular Physics, Yerevan State University,
 1 Al. Manougian str., Yerevan 375025, Armenia
 \and Ali Naji
 \at 
 %Dept. of Physics 
 Materials Research Laboratory, \& Dept. of Chemistry and Biochemistry, \& Kavli Institute for Theoretical Physics, 
 University of California, Santa Barbara, CA 93106, USA;  
 \email{anaji@chem.ucsb.edu} % (Corresponding author)
 \and Rudolf Podgornik
 \at 
 Dept. of Physics, Faculty of Mathematics and Physics, 
 University of Ljubljana and  Dept.  of Theoretical Physics, 
 J. Stefan Institute, SI--1000 Ljubljana, Slovenia, 
 \& Lab. of Physical and Structural Biology, 
 National Institutes of Health, MD 20892, USA,   %-0924
\& Kavli Institute for Theoretical Physics, University of California,
Santa Barbara, CA 93106, USA
}

% \date{\today}
\date{Received: date / Accepted: date}

\maketitle

%%%%%%%%%%%%%%%%%%%%%%%%%%%%%%%
\begin{abstract}
Charged systems with partially annealed charge disorder are investigated using field-theoretic and 
replica methods. Charge disorder is assumed to be confined to macroion surfaces surrounded
by a cloud of mobile neutralizing counterions  
in an aqueous solvent. A general formalism is developed by assuming that the 
disorder is partially annealed (with purely annealed and purely quenched
disorder included as special cases), {\em i.e.}, we assume in general that the disorder 
undergoes a slow dynamics relative to fast-relaxing 
counterions making it possible thus to study the stationary-state properties
of the system using methods similar to those available 
in equilibrium statistical mechanics. By focusing on the specific case of 
two planar surfaces of equal mean surface charge and disorder variance, it 
is shown that partial annealing of the quenched disorder 
leads to renormalization of the mean surface charge density and thus a reduction of the 
inter-plate repulsion on the mean-field or weak-coupling level.  
In the strong-coupling limit, charge disorder induces a long-range attraction resulting in  
a continuous disorder-driven collapse transition for the two surfaces 
as the disorder variance exceeds a threshold value. 
Disorder annealing further enhances the attraction and, 
in the limit of low screening, leads to a global attractive instability in the system.  
%\pacs{82.70.-y}{Disperse systems; complex fluids}
%\pacs{61.20.Qg}{Structure of associated liquids: electrolytes, molten salts, etc.}
%\pacs{87.16.D-}{Membranes, bilayers, and vesicles}
\end{abstract}
%%%%%%%%%%%%%%%%%%%%%%%%%%%%%%%

%%%%%%%%%%%%%%%%%%%%%%%%%%%%%%%%%%%%%%%%%%%%%%%%%%%%%%%%%%%%%%
\section{Introduction}

Interaction of charged macromolecules (macroions) is essential for soft and biological materials in order to maintain 
their complex structure and distinct functioning. In many cases, charge patterns 
along macromolecular surfaces are inhomogeneous and exhibit a highly disordered spatial distribution. 
DNA microarrays \cite{science95}, surfactant-coated surfaces \cite{Klein}, 
random polyelectrolytes and polyampholytes \cite{rand_polyelec} 
present examples of such disordered charge distributions. 
The charge pattern can be either set and quenched in the process of assembly of these surfaces, 
or can exhibit various degrees of annealing when interacting with other macromolecules in aqueous solutions. 
%Often the surface charges not only show a disordered distribution in space, but also
%exhibit random fluctuations in time. 
Disorder annealing in charged systems may result from 
different sources; {\em e.g.}, finite mobility and mixing of charged units (lipids and proteins) 
in lipid membranes \cite{lipowsky}, conformational rearrangement of DNA chains in 
DNA microarrays \cite{science95} and charge regulation of contact surfaces bearing weak 
acidic groups in aqueous solutions \cite{ParsegianChan}, to name a few,  all lead to annealing effects.
In reality, one may deal with a more complex situation where the surface charge pattern
displays an intermediate character  \cite{Klein} and thus may neither be considered as 
purely quenched ({\em i.e.}, with fixed random spatial distribution) nor as purely annealed ({\em i.e.}, 
thermally equilibrated with the bulk solution).

Charge disorder appears to produce electrostatic features that are remarkably different from 
those found in  non-disordered systems. Mounting experimental evidence shows that like-charged 
phospholipid membranes and fluid vesicles, which primarily contain mobile surface charges, 
may undergo aggregation and fusion in the presence of multivalent cations \cite{lipowsky}.
A similar behavior is observed with negatively charged mica surfaces exposed to a solution of
positively charged surfactants \cite{Klein}; here formation of a random mosaic of surfactant 
patches on apposing surfaces (after the surfactant is adsorbed from the bathing solution onto the surfaces)  
leads to a long-range attraction and thus a spontaneous jump to a collapsed state.
Although this behavior is akin to the transition to the primary minimum 
%when the system settles at very small inter-surface separations. 
within the standard DLVO theory of weakly charged systems \cite{DLVO,Israelachvili}, the attractive forces at work here exceed 
the universal van-der-Waals forces \cite{parsegian} incorporated in the DLVO theory by a few
orders of magnitude \cite{Klein}.  
%Such spontaneous collapse can not even be brought about solely by the recently established electrostatic strong-coupling effects that arise, {\em e.g.}, in the presence of multivalent counterions 
%such as exists in the condensation of DNA in multivalent salt solutions \cite{SC_review}. 

In fact, the emergence of an instability is not captured by the standard theories of charged systems that 
incorporate static, non-disordered charge distributions for macromolecular surfaces. These theories 
cover both mean-field  \cite{DLVO,Israelachvili,Andelman} 
or weak-coupling limit (including the Gaussian-fluctuations correction around the
mean-field solution \cite{podgornik-FI}) as well as the 
strong-coupling (SC) limit \cite{SC_review,netz,shkl,levin}, where the central theme is the absence or emergence of 
electrostatic correlations induced by neutralizing counterions in the system 
that give rise to attractive interactions between like-charged objects. 
Both uniform \cite{DLVO,Israelachvili,Andelman,netz} as well as 
modulated \cite{charge-modulation} charge distributions have been considered in this context.  In the mean-field regime, 
% (realized, {\em e.g.}, with low valency counterions),  
like-charged objects always repel. While the opposite limit of strong coupling (realized, {\em e.g.}, 
with high valency counterions, highly charged macroions, low medium dielectric constant or low temperature 
\cite{SC_review,netz}) is dominated by correlation-induced attractive forces that can bring two apposing 
like-charged surfaces to very small separation distances.  At small separations, however, a universal repulsion due to 
the confinement entropy of intervening counterions sets in and {\em stabilizes} the system in a closely packed bound state 
with a finite surface-surface separation \cite{SC_review,netz}. 
This is true even for surfaces of opposite (unequal) uniform charge distribution 
\cite{to-be-published}. Therefore, other mechanisms have to be at work that would lead to attractions strong enough to 
 counteract such repulsive forces and lead to collapse or instability in a system of 
 charged macroions. 
One such mechanism we propose is the disorder of the charge distribution along 
macromolecular surfaces that turns out to be as significant as the counterionic 
correlations and could provide a new paradigm in the theory of charged soft matter. 
%In what follows, supplementing and complementing the previous works that deal mostly with smeared 
%and static charge distributions, we shall try to establish the nature 
%and role of disorder-induced effects. Such effects turn out to be as significant as the counterionic 
%correlations and could provide a new paradigm in the theory of charged soft matter. 

Previous studies of charge disorder on macromolecular surfaces 
have investigated both types of quenched \cite{rand_polyelec,naji_podgornik,Rudi-Ali,Fleck} 
and annealed \cite{rand_polyelec,Fleck,Shklovskii_mobile,Wurger,vonGrunberg,Harries} disorder (including, specifically,
the classical work on charge-regulating surfaces \cite{ParsegianChan}). They mainly deal with situations 
where the system is in equilibrium and, on the question of electrostatic interaction \cite{Wurger,vonGrunberg,Harries}, focus
primarily on the weak-coupling regime, where disordered surfaces of equal mean 
charge always repel and no collapse or instability arises 
\cite{note_instability}. 
A systematic analysis of quenched disorder effects is  presented in the previous works of two of the present 
authors \cite{naji_podgornik,Rudi-Ali}, where it was shown that 
not only can electrostatic interactions between like-charged objects turn from repulsive to attractive due to counterionic  correlations, 
but also the disorder of the surface charge itself can give rise to an additive long-range attraction. 
This is most clearly demonstrated by attraction induced between disordered surfaces of  {\em zero} mean charge 
but with a finite variance of the disordered charge distribution  \cite{Rudi-Ali}. In the SC limit, the quenched disorder-induced 
attraction may be so strong that it can dominate the entropic repulsion at small separations and continuously shrink the SC surface-surface bound state \cite{SC_review,netz} 
upon increasing the quenched disorder variance, predicting thus a {\em continuous 
collapse transition} between a stable and a collapsed phase  beyond a threshold 
 disorder variance \cite{naji_podgornik}. Since the experimental situation may be more 
complex \cite{Klein},  and it might not allow for straightforward differentiation of the charge pattern into   
%partially quenched and partially annealed cases, 
purely quenched and purely annealed cases, 
we next set ourselves to explore possible effects from partial annealing of the surface charge. 
If there is a fingerprint of the partially annealed surface charge disorder on the nature and 
magnitude of surface interactions, this 
would help in assessing whether the experimentally observed interactions can be interpreted 
in terms of disorder-induced interactions or not. This is the motivation with which we venture on this exploration.

In this paper, we present a general formalism for charged systems 
with partially annealed disorder by invoking field-theoretic and replica methods. We
then focus on the case of two interacting {\em planar} charged surfaces as a model system and examine 
explicitly the effects of disorder on the inter-surface interaction in this system. 
Partially annealed disorder in general arises when a coupled motion of slow and fast variables (corresponding
here to surface charges and counterions, respectively) 
is present in the system. It represents a non-equilibrium situation, whose investigation requires suitable 
methods.  The previously studied cases of static, non-disordered surface charge distribution 
\cite{SC_review,netz} as well as quenched \cite{naji_podgornik} and annealed surface charge distribution 
follow as special cases from our formalism. In the SC limit, we find that 
the system of two like-charged planar surfaces with neutralizing counterions becomes 
{\em globally unstable} upon annealing the quenched surface charge
and collapses into contact regardless of other system parameters due to strong attractive forces from 
the annealing effects. Hence, the quenched phase behavior  is not stable against
small annealing perturbations  and is dramatically changed. However, stability may be restored in this system by adding a finite 
amount of added salt that screens out the long-range Coulomb interactions. In this case, we recover the continuous collapse transition 
between a stabilized  closely packed bound state of the two surfaces  and a collapsed state where the surfaces are in contact. 
This is qualitatively similar to the purely quenched case  \cite{naji_podgornik}.  
%with counterions only
However, the partially annealed bound state shows  a significantly larger attraction and a smaller optimal surface 
separation as compared to the quenched case. In other words, allowing for  rearrangements of the macroion charges leads to configurations of  
lower free energy. Since the present formalism is quite general, we shall also study the mean-field limit, where (in contrast to the 
quenched case where no disorder effects are found \cite{naji_podgornik,Fleck,note_instability}) the disorder annealing appears to suppress the mean-field 
repulsion significantly by renormalizing the surface
charge to smaller values.   Hence, besides the previously established mechanisms of counterion-induced \cite{netz,shkl,levin} 
and quenched disorder-induced \cite{naji_podgornik,Rudi-Ali}  correlations, we find that the annealing of macroion charges 
provides another mechanism enhancing the like-charge attraction. 

The organization of the paper is as follows: We start with the general formalism that allows us to define and to deal with 
the partially annealed disorder in terms of an ``effective partition function" that is obtained in the form of a functional 
integral over a fluctuating local electric potential field.  The structure of this field theory is too complicated to allow for a 
general solution. We thus derive asymptotic solutions in  
the mean-field limit (corresponding to the Poisson-Boltzmann theory of the classical DLVO framework) as well as the 
strong-coupling limit {\em via} an application of the replica formalism. We finally evaluate and analyze the inter-surface 
interactions for the specific case of planar charged surfaces  in both limits and compare them. We conclude by positioning our 
results in the growing framework of the weak--strong coupling formalism for charged macromolecular interactions.

%%%------Fig schematic 
\begin{figure}[t!]
\begin{center}
\includegraphics[angle=0,width=10cm]{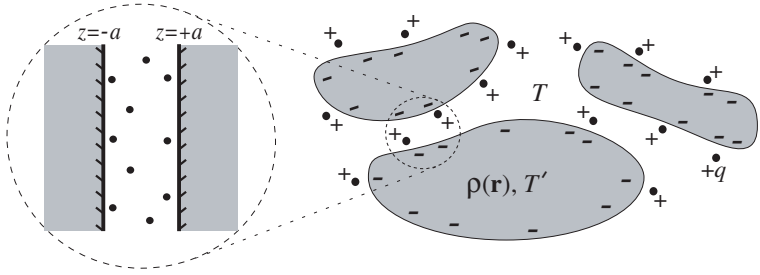}
\caption{Schematic view of a system of macroions with partially annealed 
disordered charge distribution $\rho({\mathbf r})$ and  $q$-valency 
counterions at bulk temperature $T$.  Surface charges may exhibit a 
different effective temperature $T'$ due to their disordered nature 
and slow dynamics relative to the fast-relaxing counterions. 
As a model system, we study two apposed  planar macroion surfaces with disordered 
surface charge distributions (specified in the text) located at
$z=\pm a$ at the separation distance $d=2a$. We neglect the dielectric discontinuity  
at the boundaries or the presence of added salt in the system \cite{Rudi-Ali} (see also Refs. \cite{kanduc,jho-prl,olli}). }
\label{fig:schematic}
\end{center}
%\vspace{-.5cm}
\end{figure}

%%%%%%%%%%%%%%%%%%%%%%%%%%%%%%%%%%%%%%%%%%%%%%%%%%%%%%%%%%%%%%
%%%%%%%%%%%%%%%%%%%%%%%%%%%%%%%%%%%%%%%%%%%%%%%%%%%%%%%%%%%%%%
\section{General formalism}

Let us consider a system of {\em fixed} macroions with disordered  charge distribution, $\rho({\mathbf r})$,  
immersed in an aqueous medium of 
dielectric constant, $\varepsilon$, along with their point-like neutralizing counterions of valency $q$. 
In what follows, we shall develop our formalism for an arbitrary ensemble of fixed macroions but 
for explicit calculations, we shall 
delimit ourselves to a model system of two apposed charged planar surfaces  with $\rho({\mathbf r})$ 
representing the charge distribution along both planar surfaces  (see Fig. \ref{fig:schematic}).
% that allows for a most transparent development of our approach. 
In the quenched limit, $\rho({\mathbf r})$ is assumed to be static and only counterions are subject 
to thermal fluctuations. In the annealed limit, 
both counterions and macroion charges  are subject to fluctuations of comparable time 
scales ({\em i.e.},   $\tau_{\mathrm{ci}}\sim \tau_{\mathrm{s}}$ respectively) and thus mutually 
equilibrate. The intermediate situation of partially annealed disorder by definition  
occurs when there is a macroscopic separation of time scales 
between the so-called fast and slow variables as frequently observed in glassy systems \cite{dotsenko1,dotsenko2}. 

In the present context, counterions comprise the fast variables as they are dispersed and freely fluctuate in the bulk. 
Macroion surface charges are assumed to constitute the slow variables with the time scale 
$\tau_{\mathrm{s}}\gg \tau_{\mathrm{ci}}$ due to their disordered nature (as, {\em e.g.}, they 
are confined typically within closely packed or 
quasi-two-dimensional disordered  regions such as in 
lipid bilayers \cite{lipowsky}, surfactant-coated surfaces 
or surface hemimicelles \cite{Klein,hemimicelles}).

Under these conditions, the mutual equilibration of fast and slow variables is hindered. Counterions 
rapidly attain their equilibrium at bulk temperature $T$ and thus, because of the wide time-scale gap, 
their equilibrium free energy acts as a driving force
pushing the slow dynamics of the surface charges to reach a non-equilibrium stationary state at long times. 
%But as shown because of the wide time-scale gap, it may be possible  that the slow dynamics of surface charges 
%lead to a stationary regime at long (but not infinite) times
% The whole system will of course be out of equilibrium (or at ``partial equilibrium" \cite{cool}),  and the true thermodynamic 
% equilibrium is expected to be restored gradually and at infinite times.
This scheme, known generally as the adiabatic elimination of fast variables \cite{haken,Risken,Kaneko}, 
is investigated in a growing number of works, for instance, in the context of far-from-equilibrium
stationary states and 
 thermodynamics of two-temperature systems \cite{landauer,two-T}. It has been applied in particular 
to study spin glasses with partially annealed disorder of the spin-spin coupling strength \cite{dotsenko1,dotsenko2,cool}. It has 
been shown in general that the stationary state of such systems may be described by a Boltzmann-type probability distribution 
featuring the temperature of fast degrees of freedom $T$ as well as an effective temperature $T'$
associated with the disorder. 
This peculiar two-temperature representation clearly reflects the intrinsically non-equilibrium nature of partial annealing.
%(the purely equilibrium annealed case may be reproduced as a special case with $T=T'$.)
Obviously, the equilibrium free energy of fast variables ({\em i.e.}, counterions in our case)
will show up explicitly in the aforementioned probability distribution  (here we shall not consider   
 the relaxational dynamics of the system and focus only on stationary-state properties).  

In general, one may thus define an effective
 ``partition function", ${\mathcal Z}$, in analogy with  
 the  equilibrium partition function that greatly facilitates the analysis of the 
 system far from equilibrium \cite{dotsenko1,dotsenko2,two-T,cool}.  
  This procedure is discussed in Appendix \ref{app:A} by 
 adopting  a simple dynamical model for a charged system with surface charge disorder
 and by identifying its stationary-state probability distribution. We thus find
\begin{equation}
{\mathcal Z} =  \int  {\mathcal D}\rho \, \exp\big(- \beta'  {\mathcal W}[\rho]\big),
\label{part_fun}
\end{equation}
where $\beta' = \frac{1}{k_{\mathrm{B}}T'}$ and the density functional $ {\mathcal W}[\rho]$ can be cast into the form
\begin{equation}
 \beta' {\mathcal W}[\rho] = \frac{1}{2} \int {\mathrm{d}} {\mathbf r}\,  g^{-1}({\mathbf r})\,
   								[\rho({\mathbf r}) - \rho_0({\mathbf r}) ]^2
 				- n \, \ln {\mathcal Z}_{\mathrm{ci}}[\rho],
				\label{eq:eff_pot}
\end{equation}
where 
\begin{equation}
n = \frac{T}{T'}.
\end{equation}
Equation (\ref{eq:eff_pot})  includes statistics of both counterions and  the disordered charges on 
macroion surfaces.  The first term is the contribution of the disorder. It can be interpreted 
as a general effective disorder potential expanded to the second order around a typical value $\rho_0$
(Appendix \ref{app:A}), 
which is always possible if one interprets $g({\mathbf r})$ as playing the role of 
an effective disorder ``compressibility". This expansion leads to the standard Gaussian disorder weight  with the mean value
$\rho_0({\mathbf r})$ and variance $g({\mathbf r})$ that can be handled most conveniently by replica techniques \cite{dotsenko1,dotsenko2}.
Namely, 
\begin{equation}
 {\mathcal Z}  =  \int {\mathcal D}\rho \, {\mathcal P}[\rho]\,\bigg(  {\mathcal Z}_{\mathrm{ci}}[\rho]  \bigg)^n
=  \bigg\langle   \!\!\!\!\! \bigg\langle   \bigg( {\mathcal Z}_{\mathrm{ci}}[\rho]  \bigg)^n  \bigg\rangle   \!\!\!\!\! \bigg\rangle, 
  \label{eq:Z_Zci_n}
  \end{equation}
where double-brackets denote the average $ \langle \!\! \langle  \cdots \rangle   \!\! \rangle = 
\int {\mathcal D}\rho \, {\mathcal P}[\rho]\, \big(\cdots\big)$ with respect to the Gaussian 
probability distribution 
\begin{equation}
   {\mathcal P}[\rho] = C\, \exp\bigg(\!\! - \frac{1}{2}  \int {\mathrm{d}} {\mathbf r}\,  g^{-1}({\mathbf r})\,
   								[\rho({\mathbf r}) - \rho_0({\mathbf r}) ]^2 \bigg)
\label{eq:gaussian}
\end{equation}
with $C$ being a normalization factor. 

The second term in Eq. (\ref{eq:eff_pot}) is the equilibrium free energy of a system of counterions
at a {\em fixed} realization of disordered macroion charge, $\rho = \rho({\mathbf r})$. 
It follows by integrating over the counterionic degrees of freedom equilibrated 
at temperature $T$.  
In grand-canonical ensemble, the fixed-$\rho$ partition function, ${\mathcal Z}_{\mathrm{ci}}[\rho]$, 
can be cast into a form of a functional integral as \cite{podgornik-FI,netz} 
\begin{equation}
{\mathcal Z}_{\mathrm{ci}}[\rho]  =  
				\int \frac{{\mathcal D}\phi}{{\mathcal Z}_v} 
				\,\, \, e^{- \beta  {\mathcal H}[\phi, \rho]},
\label{eq:Z_ci}
\end{equation}
where $\phi({\mathbf r})$ is the fluctuating electrostatic potential field, $\beta =\frac{1}{k_{\mathrm{B}}T}$ and 
\begin{equation}
\label{eq:H}
{\mathcal H} =  \int  {\mathrm{d}} {\mathbf r}\,
  					\bigg[ \frac{\varepsilon\varepsilon_0}{2} \big(\nabla \phi\big)^2
					             + {\mathrm{i}}\, \rho \,\phi
					- \lambda k_{\mathrm{B}}T\, 
						\Omega({\mathbf r})\, \, e^{- {\mathrm{i}}  \beta q e_0 \phi}\bigg]
\end{equation}
is the effective Hamiltonian of the system comprising Coulomb interaction  $v({\mathbf x})=(4\pi \varepsilon\varepsilon_0 |{\mathbf x}|)^{-1}$ between all charged units (the first two terms) as well as the entropy of counterions (the last term).  Here $\lambda$ is the fugacity, ${\mathcal Z}_v = \sqrt{\det \beta v({\mathbf r}, {\mathbf r}') }$,  
and $\Omega({\mathbf r})$ is a geometry function that specifies the free volume available to counterions, {\em i.e.}, the space between the two apposed planar surfaces in the model system in Fig. \ref{fig:schematic}.

The partition function (\ref{part_fun}) can be evaluated by using the replica trick \cite{dotsenko1,dotsenko2}, 
{\em i.e.}, by taking $n$ an integral number and then standardly extending the results to real axis (for any real value of $n=\beta'/\beta$) by analytical continuation. Thus by using Eq. (\ref{eq:Z_ci}) and  averaging over 
$\rho({\mathbf r})$,  we arrive at the disorder-averaged expression
\begin{equation}
\label{part_fn}   
	{\mathcal Z} = \int 
		     	\bigg(\prod_{a=1}^n \frac{{\mathcal D} \phi_a}{{\mathcal Z}_v} \bigg) \,\, \, e^{-{\mathcal S}[\{\phi_a\}]  },
\end{equation}
where the $n$-replica effective Hamiltonian reads
\begin{eqnarray}
\label{ham}
\lefteqn{
	{\mathcal S}[\{\phi_a\}] =
	\frac{1}{2}\sum_{a,b}\int {\mathrm{d}}{\mathbf r}\, {\mathrm{d}}{\mathbf r}'  \, \phi_a({\mathbf r}) {\mathcal D}_{ab}({\mathbf r}, {\mathbf r}') \phi_b({\mathbf r}') \, +
	}\\
	& & +\sum_a\int  {\mathrm{d}}{\mathbf r} \, \bigg[{\mathrm i}\beta \, \rho_0({\mathbf r}) \phi_a({\mathbf r}) -{\lambda}\Omega({\mathbf r}) \, e^{-{\mathrm i}\beta q e_0 \phi_a({\mathbf r})}\bigg].
				\nonumber
\end{eqnarray}
The kernel  ${\mathcal D}_{ab}({\mathbf r}, {\mathbf r}')$ introduced above is defined as
\begin{equation}
	{\mathcal D}_{ab}({\mathbf r}, {\mathbf r}') =
		\beta v^{-1}({\mathbf r}, {\mathbf r}') \,\delta_{ab} +
			\beta^2 g({\mathbf r})\, \delta({\mathbf r} - {\mathbf r}'), 
\label{eq:D_ab}
\end{equation}
where $v^{-1}({\mathbf r}, {\mathbf r}') = -\varepsilon\varepsilon_0 \nabla^2\delta({\mathbf r} - {\mathbf r}')$. 

Equation  (\ref{part_fn}) carries complete information about the mutual coupling between counterions and the surface charge disorder. 
The grand-canonical ``free energy" of the partially annealed system is then obtained as 
\begin{equation}
{\mathcal F} =  - k_{\mathrm B}T' \ln {\mathcal Z}.
\label{eq:free_Z}
\end{equation}
The special cases of {\em purely quenched} and {\em purely annealed} disorder follow from Eq. (\ref{eq:free_Z}) 
for $n\rightarrow 0$ and $n=1$,  respectively (see Appendix \ref{app:B}). 
%(the latter represents the fully equilibrated system with $T=T'$). 

Note that here the number of replicas, $n=T/T'$, has a direct physical meaning of temperature ratio \cite{dotsenko1,dotsenko2,two-T,cool}. A close examination of Eq.  (\ref{ham}) indicates that the partially annealed disorder gives rise to quadratic surface terms of the form $g({\mathbf r}) \phi_a({\mathbf r}) \phi_b({\mathbf r})$. It may thus 
lead to renormalization of the mean surface charge (Appendix \ref{app:eff_charge})
 as can be seen most clearly by looking at the  mean-field equations which we shall derive next.

%%%%%%%%%%%%%%%%%%%%%%%%%%%%%%%%%%%%%%%%%%%%%%%%%%%%%%%%%%%%%%
%%%%%%%%%%%%%%%%%%%%%%%%%%%%%%%%%%%%%%%%%%%%%%%%%%%%%%%%%%%%%%
\section{Mean-field limit}
\label{sec:PB}

The mean-field or Poisson-Boltzmann (PB) equation \cite{DLVO,Israelachvili,Andelman} 
(which becomes exact in the limit of small coupling parameters 
corresponding, for instance, to low counterion valency or low surface charge density \cite{SC_review,netz}) 
follows from the saddle-point equation of the functional integral (\ref{part_fn}) as
\begin{equation}  
\varepsilon \varepsilon_0 \nabla^2 \bar \phi_a =
		{\mathrm i}\,{\lambda}qe_0\,\Omega({\mathbf r})\, e^{-{\mathrm i}\beta q e_0 \bar \phi_a({\mathbf r})} +
		 {\mathrm i}\, \rho_0({\mathbf r}) + \beta g({\mathbf r})\sum_b \bar \phi_b({\mathbf r}).  
		 \end{equation}
We shall assume no preferences among different replicas on the saddle-point level, thus 
 $\bar \phi_a({\mathbf r}) = \bar \phi({\mathbf r})$ for $a = 1,\ldots, n$ (replica symmetry {\em ansatz}). In this way we arrive at  the PB equation for the real-valued mean-field potential  $ \varphi_{\mathrm{PB}}({\mathbf r})={\mathrm i} \bar \phi({\mathbf r})$ as 
\begin{equation}  
\label{eq:pb_eqn}
 \varepsilon\varepsilon_0 \nabla^2 \varphi_{\mathrm{PB}}({\mathbf r})  
		 = - {\lambda}qe_0\,\Omega({\mathbf r})\,e^{-\beta q e_0 \varphi_{\mathrm{PB}}({\mathbf r})}  - \rho_{\mathrm{eff}}({\mathbf r}), 
\end{equation}
where 
\begin{equation}
\rho_{\mathrm{eff}}({\mathbf r}) \equiv \rho_0({\mathbf r}) - \beta'  g({\mathbf r})\varphi_{\mathrm{PB}}({\mathbf r})
\label{eq:PB_rho_eff}
 \end{equation}
 is the effective (renormalized) macroion charge distribution (Appendix \ref{app:eff_charge}).  It is therefore seen that in the quenched limit ($n$ or $\beta'\rightarrow 0$), the disorder effects completely vanish on the mean-field level  
 and the PB theory coincides {\em exactly} with that of a non-disordered   
 system of bare charge distribution $\rho_0({\mathbf r})$ \cite{naji_podgornik}. 
 This is however not  true for the partially annealed disorder ($n>0$). 

To proceed with the PB theory, we shall consider the specific case of two parallel  charged plates located (normal to $z$ axis) at $z=-a$ and $z=+a$ at the separation distance $d=2a$ (Fig. \ref{fig:schematic}). 
We take the mean charge distribution and its variance as 
\begin{eqnarray}
	\rho_0({\mathbf r}) &=& -\sigma e_0 \,\big[\delta(z+a) +\, \delta(z-a) \big] 
	\label{eq:sigma_plates} \\
	g({\mathbf r}) &=& g \,e_0^2 \,\big[\delta(z+a) + \delta(z-a) \big], 
	\label{eq:g_plates}
\end{eqnarray}
where $g\geq 0$ and without loss of generality we assume that $\sigma\geq 0$ (and thus $q\geq 0$).  
Counterions are assumed to be confined in between the plates ({\em i.e.}, $\Omega({\mathbf r}) = 1$ for $|z|<a$ 
and zero elsewhere), where Eq. (\ref{eq:pb_eqn}) admits the well-known solution \cite{DLVO,Israelachvili,Andelman}
\begin{equation} 
\varphi_{\mathrm{PB}}(z) = \frac{1}{\beta q e_0}\ln \cos^2(K z )
\end{equation} 
with $K^2=2\pi \ell_{\mathrm{B}} q^2  {\lambda}$  to be determined from the electroneutrality condition stipulating that the total charge on the two surfaces should be equal to the total charge of the counterions. This leads to the equation for $K$ of the form 
\begin{equation}
K\mu\, \tan(Ka) = 1 + \gamma \ln \cos^2(K a) \equiv \frac{\sigma_{\mathrm{eff}}}{\sigma}.
\label{eq:PB_K}
\end{equation}
Here $\mu=1/(2\pi q \ell_{\mathrm{B}} \sigma)$ is the Gouy-Chapman length, $\ell_{\mathrm{B}} = e_0^2/(4\pi \varepsilon \varepsilon_0 k_{\mathrm{B}}T)$ the Bjerrum length, and 
\begin{equation} 
\sigma_{\mathrm{eff}} = \sigma + \beta' g e_0 \varphi_{\mathrm{PB}}(a)
\label{renormedsigma}
\end{equation}
the renormalized surface charge density (Eq. (\ref{eq:PB_rho_eff})). The latter expression 
clearly reflects the  mixed boundary conditions encountered here, resembling 
the situation in the classical charge-regulation problems \cite{ParsegianChan}.  The dimensionless parameter
\begin{equation}
\gamma = \frac{n g}{q \sigma}
\label{eq:gamma}
\end{equation}
gives a measure of the {\em disorder annealing} and is obviously proportional to the ratio $n=T/T'$ of the counterions and surface disorder temperatures. 

%%%------Fig two plates PB pressure
\begin{figure}[t!]
\begin{center}
\includegraphics[angle=0,width=7.5cm]{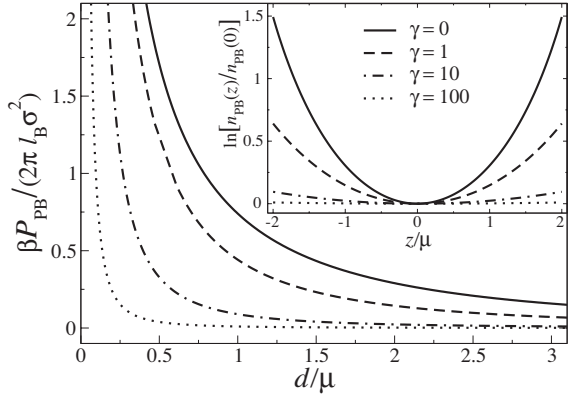}
\caption{Rescaled PB pressure between two charged plates as a function of their separation $d$ for $\gamma = 0$, 1, 10 and $10^2$. Inset shows the PB counterion density profile for $d/\mu=4$. For $\gamma=0$, we recover the non-disordered results with the pressure decaying as $\sim 1/d^2$ \cite{netz}. 
For $\gamma\gg 1$,  the pressure decays as $\sim 1/(\gamma d^2)$. }
\label{fig:pb_pressure_density}
\end{center}
%\vspace{-.5cm}
\end{figure}

The PB pressure, $P_{\mathrm{PB}}$, acting between the plates is obtained from the standard definition 
$\beta P_{\mathrm{PB}} = n_{\mathrm{PB}}(z_0) -  \frac{1}{2}\beta\varepsilon\varepsilon_0 ({\mathrm{d}}\varphi_{\mathrm{PB}}/{\mathrm{d}} z)^2|_{z_0}$  \cite{DLVO,Israelachvili,Andelman} for an arbitrary $|z_0|<a$ as
\begin{equation} 
\frac{\beta P_{\mathrm{PB}}}{2\pi \ell_{\mathrm{B}} \sigma^2} = (K\mu)^2. 
\end{equation}
The counterion number density profile between the plates, $n_{\mathrm{PB}}(z) = {\lambda}\,e^{-\beta q e_0 \varphi_{\mathrm{PB}}(z)}$ \cite{DLVO,Israelachvili,Andelman}, is obtained as  
\begin{equation} 
\frac{n_{\mathrm{PB}}(z)}{2\pi \ell_{\mathrm{B}} \sigma^2}=\left(\frac{K\mu}{\cos Kz}\right)^2. 
\end{equation}  

It follows from Eq. (\ref{eq:PB_K}) that the mean-field renormalized surface charge density is always smaller than the bare value 
and tends to zero but never changes sign as $\gamma$  increase ($0\leq \sigma_{\mathrm{eff}} \leq \sigma$).  
Therefore, the surfaces are effectively neutralized and the pressure as well as  the counterion number density profile  
tend to zero as $\gamma$  increases to infinity (see Fig. \ref{fig:pb_pressure_density}). 
This picture relies on the assumption that the number of
surface charged units is not fixed and can respond to changes of the surface potential. Imposing
the constraint that fixes this number obviously rules out surface charge renormalization and  
one observes no effects from the disorder annealing in agreement with Ref. \cite{Fleck}. 

In the limit $\gamma \rightarrow 0$, we recover the non-disordered \cite{netz} or quenched  \cite{naji_podgornik}
mean-field results with the following asymptotic behavior for the pressure, 
\begin{equation} 
\frac{\beta P_{\mathrm{PB}}}{2\pi \ell_{\mathrm{B}} \sigma^2} \simeq
    \left\{
	             \begin{array}{ll}
       		       2\mu/d
             		& {\qquad d/\mu\ll 1,}\\	\\	            
		        \pi^2\mu^2/d^2
	              & {\qquad d/\mu \gg 1}.
     \end{array}
       \right.  
\end{equation}
 For  large $\gamma\gg 1$,  we find that $(Ka)^2\simeq (\mu/a+\gamma)^{-1}$ and thus
\begin{equation} 
\frac{\beta P_{\mathrm{PB}}}{2\pi \ell_{\mathrm{B}} \sigma^2} \simeq
    \left\{
	             \begin{array}{ll}
       		       2\mu/d
             		& {\qquad \gamma d/\mu\ll 1,}\\	\\	            
		        4\mu^2/(\gamma d^2)
	              & {\qquad \gamma d/\mu \gg 1}.
     \end{array}
       \right.  
\end{equation}
The small separation expression above is nothing but the ideal-gas osmotic pressure of counterions 
that dominates over the energetic contributions. At large separations the pressure is found to 
decay asymptotically as $\sim 1/(\gamma d^2)$.  
The pressure remains always non-negative and the surface-surface interaction is thus always repulsive in 
the mean-field limit.  Choosing the non-disordered system as the reference, however, 
the decrease in the interaction pressure upon increase of the surface disorder annealing 
can be interpreted as being due to an effective disorder-induced attraction whose asymptotic form 
could be described by 
\begin{equation}
\Delta P_{\mathrm{PB} }\sim \frac{1 - \gamma }{\gamma} \frac{1}{d^2}
\end{equation}
for large $\gamma$.  
This asymptotic form again attests to the fact that the way the disorder acts on the mean-field 
interaction between the two apposed surfaces is {\em via} a renormalization of the surface charge density.

%%%%%%%%%%%%%%%%%%%%%%%%%%%%%%%%%%%%%%%%%%%%%%%%%%%%%%%%%
%%%%%%%%%%%%%%%%%%%%%%%%%%%%%%%%%%%%%%%%%%%%%%%%%%%%%%%%%
\section{Strong-coupling (SC) limit}

Next we shall investigate the asymptotic strong-coupling limit which is complementary to the mean-field
limit and where counterion-induced correlations become dominant. 
We employ the standard strong-coupling scheme reviewed extensively in Refs.  \cite{SC_review,netz}  in order 
to study the partial annealing effects in the SC limit. The so-called asymptotic  
SC theory is obtained from the leading order terms of a non-trivial virial expansion
(in powers of the fugacity) of the partition function (\ref{part_fn}), {\em i.e.} 
\begin{equation} 
{\mathcal Z} = {\mathcal Z}_0 + \lambda {\mathcal Z}_1 + {\mathcal O}({\lambda}^2). 
\label{virialterms}
\end{equation} 
It becomes exact in the limit of large coupling parameters corresponding, 
for instance, to high counterion valency or high surface charge density  \cite{SC_review,netz}. 

The zeroth-order (no counterion) term, ${\mathcal Z}_0$, and the first-order (single counterion) term, 
${\mathcal Z}_1$, follow from Eq. (\ref{part_fn}) as 
\begin{eqnarray}
	{\mathcal Z}_0 = \int\bigg(\prod_{a=1}^n \frac{{\mathcal D} \phi_a}{{\mathcal Z}_v} \bigg) \,\, e^{-{\mathcal S}_0},
	\label{eq:Z_0}
\\
	\label{eq:Z_1}
	{\mathcal Z}_1 = \sum_{b=1}^n\int  {\mathrm{d}}{\mathbf R} \, \Omega({\mathbf R})
		     	\int\bigg(\prod_{a=1}^n \frac{{\mathcal D} \phi_a}{{\mathcal Z}_v} \bigg) \,\, e^{-{\mathcal S}_0 - 
				{\mathrm i}\beta q e_0  \phi_b({\mathbf R}) },
\end{eqnarray}
where  
\begin{equation} 
{\mathcal S}_0 = \frac{1}{2}\sum_{a,b} \int {\mathrm{d}}{{\mathbf r}\, {\mathrm{d}}{\mathbf r}'} \, \phi_a({\mathbf r}){\mathcal D}_{ab}({\mathbf r}, {\mathbf r}')  \phi_b({\mathbf r}')  +{\mathrm i}\beta \sum_a \int {\mathrm{d}}{\mathbf r} \, \rho_0({\mathbf r}) \phi_a({\mathbf r}).
\end{equation} 
We thus need to calculate both these terms for an arbitrary number of replicas, $n$. In doing so, we shall make use of some mathematical relations that we briefly discuss below.

\subsection{Mathematical preliminaries}
\label{subsec:math}

First, it turns out that the most convenient way to carry out the calculations is to replace the 
long-range Coulomb interaction  $v({\mathbf x})=1/(4\pi \varepsilon\varepsilon_0 |{\mathbf x}|)$ 
with the exponentially screened Yukawa interaction 
\begin{equation}
v_s({\mathbf x})=\frac{e^{-\kappa |{\mathbf x}|}}{4\pi \varepsilon\varepsilon_0 |{\mathbf x}|}
\end{equation}
 by introducing a finite screening length $\kappa^{-1}$. In the end, we shall take the limit $\kappa\rightarrow 0$. Note that not only is this procedure of technical convenience but it is also of physical
relevance for the present problem. It corresponds to adding a background salt to the system, leading to the screened Coulomb interaction between charged units. It is to be noted however that the salt effects are taken into account in this way only on the 
linear Debye-H\"uckel level. (Further study of the role of added salt in the SC limit is presented elsewhere \cite{olli}.)  
Generalization of Eqs. (\ref{part_fn})-(\ref{eq:D_ab}) in the presence of Yukawa interaction is immediate as $v^{-1}({\mathbf r}, {\mathbf r}')$ is simply replaced by  
\begin{equation}
v_s^{-1}({\mathbf r}, {\mathbf r}') = \varepsilon\varepsilon_0 (-\nabla^2+\kappa^2)\delta({\mathbf r} - {\mathbf r}').
\end{equation}

Second, in calculating ${\mathcal Z}_0$ and ${\mathcal Z}_1$ one needs to evaluate the determinant and the inverse of the block-matrix  ${\mathcal D}_{ab}({\mathbf r}, {\mathbf r}')$. These calculations are straightforward
and may be carried out most easily by employing properties of block-matrices and the operator algebra defined over the Hilbert space $\{|{\mathbf r}\rangle\}$. We shall use the compact notation by defining the operators 
$\langle {\mathbf r}|\hat v_s|{\mathbf r}'\rangle=v_s({\mathbf r}, {\mathbf r}')$, $\langle {\mathbf r}|\hat g|{\mathbf r}'\rangle=g({\mathbf r})\,\delta({\mathbf r}, {\mathbf r}')$ (for the screened Coulomb interaction and disorder variance), 
and $\langle {\mathbf r}|\hat {\mathbf D}_{ab}|{\mathbf r}'\rangle={\mathcal D}_{ab}({\mathbf r}, {\mathbf r}')$
{\em via} Eq. (\ref{eq:D_ab}), where the latter is defined as an element of the  $n\times n$ operator matrix 
\begin{equation}
\hat {\mathbf D}=\beta\, {\mathbf e}\otimes \hat v_s^{-1}+\beta^2\,{\mathbf u}\otimes \hat g
\end{equation}
 with ${\mathbf e}_{ab}=\delta_{ab}$ and ${\mathbf u}_{ab}=1$. 
Also, we shall use $\langle {\mathbf r}|\rho_0\rangle = \rho_0({\mathbf r})$ and the well-known notation 
$\int_{{\mathbf r}, {\mathbf r}'} \rho_0({\mathbf r}) v_s({\mathbf r}, {\mathbf r}') \rho_0({\mathbf r}') = 
\langle \rho_0|\hat v_s|\rho_0\rangle$, etc. 

%By induction 
One can prove the following identities for $\hat {\mathbf D}$ 
\begin{eqnarray}
  \det  \hat {\mathbf D} = \big(\det \beta \hat v_s^{-1}\big)^{n}\, \det \big(\hat 1+n \beta \hat g \hat v_s \big),
  \label{eq:det_Dab}\\
  \beta \sum_a \big\langle {\mathbf r}\big|(\hat {\mathbf D}^{-1})_{ab}\big|{\mathbf r}'\big\rangle 
  		=\big\langle {\mathbf r}\big|\hat v_s \big(\hat 1+n \beta \hat g \hat v_s\big)^{-1}\big|{\mathbf r}'\big\rangle, 
  	\label{eq:inverse_Dab}\\
  \beta \big\langle {\mathbf r}\big| (\hat {\mathbf D}^{-1})_{aa}\big|{\mathbf r}\big\rangle
  		=\big\langle {\mathbf r}\big|\hat v_s \big[\hat 1 - \beta \hat g \hat v_s
     \big(\hat 1+n \beta \hat g \hat v_s\big)^{-1}\big]\big|{\mathbf r}\big\rangle, 
     \label{eq:inverse_Daa}
\end{eqnarray}
which will be used in what follows. Note that the last two quantities do not depend on the replica indices $a$ and $b$.

\subsection{ Virial terms ${\mathcal Z}_0$ and ${\mathcal Z}_1$}
\label{subsec:virial}

Going back to the virial term ${\mathcal Z}_0$, one can perform the Gaussian integral in Eq. (\ref{eq:Z_0}) to obtain
\begin{equation}
	{\mathcal Z}_0 = C_0\,e^{-\frac{n}{2}\ln \det \beta\hat v_s -\frac{1}{2}\ln \det  \hat {\mathbf D} 
	- \frac{1}{2}\beta^2 \sum_{a,b}\langle \rho_0|(\hat {\mathbf D}^{-1})_{ab}|\rho_0\rangle}.
	\label{eq:Z_0_ii}
\end{equation}
Using Eqs. (\ref{eq:det_Dab}) and (\ref{eq:inverse_Dab}), ${\mathcal Z}_0$ is completely determined. 
This term represents the interaction free energy of macroion charges in the absence of counterions. 

Next, the Gaussian integral in Eq. (\ref{eq:Z_1}) can be evaluated as
\begin{equation}
	{\mathcal Z}_1 = n {\mathcal Z}_0 \int  {\mathrm{d}}{\mathbf R} \, \Omega({\mathbf R})
	\, e^{-\beta u({\mathbf R})}, 
\label{eq:Z_1_ii}	
\end{equation}
where $u({\mathbf R})$ is nothing but the single-counterion interaction energy with the macroion charges
and reads 
\begin{equation}
 u({\mathbf R}) = \beta q e_0 \sum_a\langle \rho_0|(\hat {\mathbf D}^{-1})_{ab}|{\mathbf R} \rangle 
 	+\beta\frac{(q e_0)^2}{2}  \langle {\mathbf R} |(\hat {\mathbf D}^{-1})_{bb}|{\mathbf R} \rangle.
\label{eq:u_general}
\end{equation}
This is fully determined by virtue of Eqs. (\ref{eq:inverse_Dab}) and (\ref{eq:inverse_Daa}). We have thus derived the  general 
form of both virial terms in Eq. (\ref{virialterms}) as a function of $n$. We now proceed to the explicit evaluation 
of the two virial terms for the system of two apposed charged planar surfaces as defined 
by the mean charge distribution and disorder variance (\ref{eq:sigma_plates})  and (\ref{eq:g_plates}) (see Fig. \ref{fig:schematic}).

\subsection{Small-$n$ expansion}
\label{subsec:small_n}

To proceed further we focus on the small-$n$ limit of the above expressions. Our chief goal here is
 to examine the stability of the system upon small annealing perturbations of a quenched charge distribution. 
 The annealing effects on this level are therefore expected to be additive in the free energy of the system.

By expanding Eq. (\ref{eq:Z_0_ii}) for small $n$ we arrive at
\begin{eqnarray}
	{\mathcal Z}_0 \simeq C_0'\,\exp\bigg(-\frac{n\beta}{2}\bigg[{\mathrm{Tr}}(\hat g\hat v_s)
	   + \langle \rho_0|\hat v_s|\rho_0\rangle\bigg] + \nonumber\\
	   +\frac{(n\beta)^2}{2}
	   \bigg[\frac{1}{2}{\mathrm{Tr}}\big(\{\hat g\hat v_s\}^2\big) 
	   + \langle \rho_0|\hat v_s \hat g \hat v_s|\rho_0\rangle\bigg] \bigg)
\label{eq:Z_0_approx}
\end{eqnarray}
to the lowest orders in $n$. But since we are interested in the inter-plate interaction, we shall
need to determine only the separation-dependent terms. 
 
It easily follows that the ${\mathrm{Tr}}(\hat g\hat v_s)$ 
term in the above equation does not depend on the inter-surface 
distance $d=2a$ and the ${\mathrm{Tr}}(\{\hat g\hat v_s\}^2)$ term 
may be calculated straightforwardly (and up to an irrelevant additive term) as 
\begin{equation}
  \beta^2\,{\mathrm{Tr}}(\{\hat g\hat v_s\}^2) = -S\, (4\pi \ell_{\mathrm{B}}^2 g^2)\, {\mathrm{Ei}}(-2\kappa d), 
  \label{eq:Trgv^2}
\end{equation} 
where $S$ is the total area  of each surface and ${\mathrm{Ei}}(x)=\int_{-\infty}^x {\mathrm{d}}t\, e^{t}/t$
is the exponential-integral function. 
The remaining two terms in the expression for ${\mathcal Z}_0$ are obtained as
\begin{eqnarray}
\label{eq:rho_v_rho}
 \beta \langle \rho_0|\hat v_s|\rho_0\rangle  = 2S\,\big(\sigma^2\ell_{\mathrm{B}} \big) \bigg(\frac{2\pi}{\kappa}\bigg)\big(1 + e^{-        	\kappa d}\big),\\
 \beta^2\langle \rho_0|\hat v_s \hat g \hat v_s|\rho_0\rangle = 
 	2S\,\big(g\sigma^2 \ell_{\mathrm{B}}^2 \big) \bigg(\frac{2\pi}{\kappa}\bigg)^2
		\big(1 + e^{-\kappa d}\big)^2.
\end{eqnarray}

The small-$n$ expansion for ${\mathcal Z}_1$ leads to evaluation of  
$e^{-\beta u({\mathbf R})}\simeq e^{-\beta u_0({\mathbf R})}\big[1+n \beta u_1({\mathbf R})\big]$, where
we have expanded $u({\mathbf R}) = u_0({\mathbf R}) - n u_1({\mathbf R}) + {\mathcal O}(n^2)$ to the lowest order in $n$, and thus 
the following two terms
\begin{eqnarray}
  u_0({\mathbf R}) = q e_0 \langle \rho_0|\hat v_s|{\mathbf R} \rangle 
 	- \beta\frac{(q e_0)^2}{2} \langle {\mathbf R} |\hat v_s \hat g\hat v_s|{\mathbf R} \rangle, 
	\label{eq:u_0}\\
  u_1({\mathbf R}) = \beta q e_0 \langle \rho_0|\hat v_s \hat g\hat v_s|{\mathbf R} \rangle 
 	- \beta^2\frac{(q e_0)^2}{2}  \langle {\mathbf R} |\hat v_s \hat g\hat v_s\hat g\hat v_s|{\mathbf R} \rangle. 
	\label{eq:u_1}
\end{eqnarray}
We have discarded a self-energy term $(qe_0)^2\langle {\mathbf R} |\hat v_s |{\mathbf R} \rangle$ 
in Eq. (\ref{eq:u_0}) whose only effect is to rescale
the fugacity, $\lambda$ \cite{naji_podgornik}. 
We can then use the explicit expressions
\begin{eqnarray}
\beta e_0 \langle \rho_0|\hat v_s|{\mathbf R} \rangle  &=& 
	-\big(\sigma\ell_{\mathrm{B}}\big)\bigg(\frac{2\pi}{\kappa}\bigg)\big(e^{-\kappa|a-R_z|}+e^{-\kappa|a+R_z|}\big) \\
\beta^2 e_0 \langle \rho_0 |\hat v_s \hat g\hat v_s|{\mathbf R} \rangle 
 	&=& - \big(\sigma g\ell_{\mathrm{B}}^2\big)\bigg(\frac{2\pi}{\kappa}\bigg)^2\big(1 + e^{-\kappa d}\big) \big(e^{-\kappa|a-R_z|}+e^{-\kappa|a+R_z|}\big). 
\end{eqnarray}
The two remaining expressions in Eqs. (\ref{eq:u_0}) and (\ref{eq:u_1}) are obtained as
\begin{eqnarray}
\beta^2 e_0^2\langle {\mathbf R} |\hat v_s \hat g\hat v_s|{\mathbf R} \rangle  &=& -\big(2\pi g \ell_{\mathrm{B}}^2\big)
	\bigg[{\mathrm{Ei}}(-2\kappa |a-R_z|) + {\mathrm{Ei}}(-2\kappa |a+R_z|)\bigg], \\
\beta^3 e_0^2\langle {\mathbf R} |\hat v_s \hat g\hat v_s\hat g\hat v_s|{\mathbf R} \rangle 
		& =& \big(2\pi g^2 \ell_{\mathrm{B}}^3 \big) 
 		\bigg(\frac{2\pi}{\kappa}\bigg) \bigg[\big(e^{-2\kappa |a-R_z|}+e^{-2\kappa |a+R_z|}+2e^{-2\kappa d}\big)+ \nonumber\\
		&& + 2\kappa |a-R_z|\,{\mathrm{Ei}}(-2\kappa |a-R_z|) + 2\kappa |a+R_z|\,{\mathrm{Ei}}(-2\kappa |a+R_z|) + \nonumber\\
		 && + 4\kappa d\,\, {\mathrm{Ei}}(-2\kappa d)\bigg]. 
\end{eqnarray}
We shall need only the small $\kappa$ results, which read (after discarding irrelevant constants)
\begin{eqnarray}
\beta^2 e_0^2\langle {\mathbf R} |\hat v_s \hat g\hat v_s|{\mathbf R} \rangle  &\simeq& -\big(2\pi g \ell_{\mathrm{B}}^2\big)
	\ln \big(a^2-R_z^2 \big), \\
\beta^3 e_0^2\langle {\mathbf R} |\hat v_s \hat g\hat v_s\hat g\hat v_s|{\mathbf R} \rangle & \simeq& \big(2\pi g^2 \ell_{\mathrm{B}}^3 \big)
 		\bigg(\frac{2\pi}{\kappa}\bigg) \big(e^{-2\kappa|a-R_z|}+e^{-2\kappa|a+R_z|}+2e^{-2\kappa d}\big).\nonumber\\
	\label{eq:R_vgvgv_R}
\end{eqnarray}
This completes the explicit evaluation of both virial terms. We now proceed to the evaluation of the interaction 
free energy, {\em i.e.}, the part of the free energy that explicitly depends on the separation between the two surfaces.

\subsection{SC free energy}

By using  Eqs. (\ref{eq:Z_0_ii})-(\ref{eq:u_general}), one can evaluate the free energy 
 of a partially annealed system from ${\mathcal F}^{\mathrm{SC}}= -k_{\mathrm{B}}T' \ln {\mathcal Z} $ 
 (Eq. (\ref{eq:free_Z})), 
 where ${\mathcal Z} =  {\mathcal Z}_0 + \lambda {\mathcal Z}_1$  in the SC limit  
 (Eq. (\ref{virialterms})).  
 The fugacity can be fixed by the number of counterions $N$ upon transforming to canonical ensemble {\em via} 
\begin{equation}
n N = \lambda \frac{\partial \ln {\mathcal Z}}{\partial \lambda},
\label{eq:N_lambda}
\end{equation}
 whereby we obtain 
\begin{equation}
\lambda = \frac{N}{\int {\mathrm{d}}{\mathbf R} \, \Omega({\mathbf R})\, e^{-\beta u({\mathbf R})}}.
\label{eq:lambda}
\end{equation}
The canonical SC free energy is obtained {\em via}  the Legendre transform, 
${\mathcal F}_N^{\mathrm{SC}} = {\mathcal F}^{\mathrm{SC}} + Nk_{\mathrm{B}}T \ln \lambda $, as
\begin{equation}
\frac{\beta{\mathcal F}_N^{\mathrm{SC}}}{N} = -\frac{\ln  {\mathcal Z}_0}{Nn} 
	- \ln  \int  {\mathrm{d}}{\mathbf R}\, \Omega({\mathbf R})\, e^{-\beta u({\mathbf R})}
\label{eq:SCfree_general}
\end{equation}
supplemented by the electroneutrality condition, which for the  two-plate system reads $N q = 2 \sigma_{\mathrm{eff}} S$, 
and again stipulates that the total charge on the surfaces equals the charge of the interposed counterions. 
Note that here $\sigma_{\mathrm{eff}} $ is the effective surface charge density that has to be determined self-consistently 
within the SC theory. It turns out however that, in the SC limit, there is no charge renormalization
due to partial annealing on the leading order in $n$, 
and thus   $\sigma_{\mathrm{eff}} = \sigma$  up to corrections of the 
order ${\mathcal O}(n^2)$ (see Appendix \ref{app:eff_charge}). 
This behavior is in stark contrast with the one found on the mean-field level in Section \ref{sec:PB}. 

The above expressions (\ref{eq:N_lambda})-(\ref{eq:SCfree_general}) together with Eqs. (\ref{eq:Z_0_ii})-(\ref{eq:u_general})
are applicable to any general system of fixed macroions with 
(Gaussian) disordered charge distribution and arbitrary degree of annealing $n$.
For small annealing perturbations, we have 
\begin{equation}
\frac{\beta{\mathcal F}_N^{\mathrm{SC}}}{N} = -\frac{\ln  {\mathcal Z}_0}{Nn} 
	- \ln  \int  {\mathrm{d}}{\mathbf R}\, \Omega({\mathbf R})\, e^{-\beta u_0({\mathbf R})} - n\beta
	  \frac{\int {\mathrm{d}}{\mathbf R} \, \Omega({\mathbf R})\, u_1({\mathbf R})\, e^{-\beta u_0({\mathbf R})} }
    					  {\int {\mathrm{d}}{\mathbf R} \, \Omega({\mathbf R})\, e^{-\beta u_0({\mathbf R})} } + {\mathcal O}(n^2), 
\end{equation}
where  $ {\mathcal Z}_0$ is given by  Eq. (\ref{eq:Z_0_approx}). 

In what follows, we shall focus again on the two-plate model system
(Eqs. (\ref{eq:sigma_plates})  and (\ref{eq:g_plates})) and make use of the explicit 
expressions (\ref{eq:Trgv^2})-(\ref{eq:R_vgvgv_R})
in order to calculate the SC free energy of this system. 
We then take the limit of small inverse screening length, $\kappa\rightarrow 0$, as  
noted in Section \ref{subsec:math}. We thus find that the SC free energy of this system 
adopts a simple form as
\begin{eqnarray}
 \frac{\beta {\mathcal F}_N^{\mathrm{SC}}}{N} &\simeq& f_{\mathrm{quenched}} + 
 		\frac{n}{\kappa} \big(8\pi^2 q \ell_{\mathrm{B}}^2 g \sigma\big) d, \nonumber\\
		&= & f_{\mathrm{quenched}} + \bigg(\frac{2\gamma}{\kappa \mu}\bigg)  \frac{d}{\mu},
	\label{eq:SCfree}	
\end{eqnarray}
where the first term on the right hand side, $f_{\mathrm{quenched}} $, 
 is nothing but the quenched ($n=0$) rescaled free energy \cite{naji_podgornik}
\begin{equation}
   f_{\mathrm{quenched}}  = \frac{d}{2\mu} + (\chi -1)\ln d,
   \label{eq:SCfree_quenched}
 \end{equation}  
and the second term in Eq. (\ref{eq:SCfree}) 
represents the leading order correction from
partial annealing of the disorder. Note that this term is
linear in $n$ as expected and scales with the inverse screening 
length $\kappa^{-1}$. 

The quenched free energy (\ref{eq:SCfree_quenched}) is expressed in terms of the dimensionless {\em disorder coupling parameter}  
$\chi = 2\pi q^2 \ell_{\mathrm{B}}^2 g$, which  gives a measure of the disorder-induced coupling {\em via} 
the surface charge  variance $g$. This is to be compared with the so-called {\em electrostatic 
coupling parameter} $\Xi = 2\pi q^3 \ell_{\mathrm{B}}^2 \sigma$ defined originally for non-disordered 
systems \cite{netz}, which measures the strength of counterion-induced correlations and, in the present 
context, depends on the mean charge density $\sigma$. The information about the  disorder annealing enters only {\em via} 
$\gamma = n g/(q \sigma)$ as defined previously in Eq. (\ref{eq:gamma}).
 These dimensionless parameters may be used to determine the phase behavior of a 
 partially annealed system, with the disorder effects being in general quantified by $\chi$ and $\gamma$.

\subsection{Instability and collapse transition}

The quenched free energy (\ref{eq:SCfree_quenched}) comprises  the standard non-disordered SC contributions,  
{\em i.e.}, the counterion-mediated attraction, $d/2\mu$, and  the  repulsion, $-\ln d$, due to the confinement entropy of counterions between the two surfaces \cite{SC_review,netz}. But it also includes a long-range additive logarithmic term, that is $\chi \ln d$, stemming from the disorder variance which is attractive and renormalizes the repulsive entropic term. 
 This peculiar  form of the quenched disorder 
 contribution leads to the previously predicted  \cite{naji_podgornik} {\em continuous collapse transition} 
 at the threshold $\chi_c=1$ between a stable bound state of the two surfaces  and a collapsed state 
 where the surfaces are in contact. In other words, the optimal surface-surface 
 separation, $d_\ast$,  behaves as
\begin{equation}
	 \frac{d_\ast}{\mu} =\left\{
	             \begin{array}{ll}
       		       2(1-\chi)
             		& {\qquad \chi < 1,}\\	\\	            
		        0
	              & {\qquad \chi > 1}.
     \end{array}
       \right.  
\end{equation}

If on the other hand the disorder is partially annealed ($n>0$), we see from Eq. (\ref{eq:SCfree})  that the annealing  generates a {\em linear} attractive term in the free energy,  that is $\sim \gamma d/\kappa$. It   adds up with and enhances the counterion-mediated attraction term 
(first term in Eq. (\ref{eq:SCfree_quenched})) 
exhibiting thus a complementary effect when compared to the quenched disorder contribution. 
The threshold of the collapse transition $\chi_c=1$  remains intact up to small 
corrections of the order ${\mathcal O}(n)$ \cite{note:threshold} and the 
partially annealed bound-state separation is obtained in the limit of small inverse screening length, $\kappa$, as
\begin{equation}
	 \frac{d_\ast}{\mu} \simeq \left\{
	             \begin{array}{ll}
       		       \frac{2(1-\chi)}{1 +4\gamma/\kappa\mu}
             		& {\qquad \chi < 1,}\\	\\	         
		        0
	              & {\qquad \chi > 1},
     \end{array}
       \right.  
       \label{eq:d*_partial}
\end{equation}
which is always smaller than the quenched value, reflecting again the enhanced surface-surface attraction in the case of partially annealed disorder. 

Note that the optimal separation in the partially annealed  case, Eq. (\ref{eq:d*_partial}), 
tends to zero, $d_\ast\rightarrow 0$, as the inverse screening length tends to zero $\kappa\rightarrow 0$.
In other words,  the system goes into a collapsed state regardless of other parameters exhibiting thus  
a {\em global attractive instability}.  Hence, a quenched system of macroion charges and counterions is not stable with respect to annealing perturbations of the macroion charge
distribution in the absence of screening effects. The stability may be achieved by adding a finite amount
of added salt. 
Unfortunately, the effects of added salt have not yet been properly 
analyzed in the context of the SC theory (see 
Ref. \cite{olli} for a recent attempt) and it is presently difficult to go beyond 
the linear description adopted here for the salt screening effects. 

%%%%%%%%%%%%%%%%%%%%%%%%%%%%%%%%%%%%%%%%%%%%%%%%%%%%%%%%%
%%%%%%%%%%%%%%%%%%%%%%%%%%%%%%%%%%%%%%%%%%%%%%%%%%%%%%%%%
\section{Conclusion and discussion} 

In this work we have analyzed the effects of partially annealed disorder in the distribution of macromolecular (macroion) charges on the interaction between two such macromolecular surfaces across a solution containing
mobile neutralizing counterions. Recent experiments on decorated mica surfaces \cite{Klein} covered with a random mosaic of positive and negative charged domains (stemming from the adsorption of cationic surfactants) 
clearly point to the existence of a strong attractive surface-surface interaction which, upon formation of such domains, 
collapses the system into a compact state with the surfaces being in contact. 
This behavior resembles the transition to a primary minimum within the DLVO theory \cite{DLVO,Israelachvili} although the 
attraction mechanism here is strictly non-DLVO  \cite{Klein} producing attractive electrostatic forces that are up to a few orders of magnitude larger than the universal van-der-Waals forces \cite{parsegian} as incorporated in the DLVO theory. 
Moreover, the emergence of an attractive  instability or even a collapse transition is not predicted within the standard theories
of electrostatic interactions between charged macromolecular surfaces (even between surfaces bearing 
opposite charges) pointing thus
to the possible role of the surface charge disorder in the aforementioned collapse transition. 

The AFM investigation of the surface texture of mica surfaces \cite{Klein}  can be performed only before the measurements of the inter-surface forces and can not be monitored while the two surfaces are brought closer together. One can thus not be sure whether the surface distribution of the charged domains along the surfaces changes on approach of the surfaces or not.  This leads in general to three possible scenarios: 
\begin{itemize}
\item the surface charge disorder is completely set by the method of preparation of the surfaces and does not change on approach of the surfaces (quenched disorder), 
\item the surface charge disorder responds to the changes in the separation of the surfaces just as fast as the mobile charged species (such as counterions) in the solution between the surfaces (annealed disorder), 
\item and lastly, the surface charge disorder does respond to the changes of the separation but with a much larger time-scale than the mobile charged species between the surfaces  (partially annealed disorder). 
\end{itemize}
As the experiment alluded to above gives only the inter-surface interaction as a function of the separation, one is thus lead to investigate the {\em fingerprint} of these different scenarios on the behavior of the interaction. Since the effects of annealed 
\cite{ParsegianChan,Fleck,Shklovskii_mobile,Wurger,vonGrunberg,Harries} and quenched \cite{naji_podgornik,Rudi-Ali,Fleck} disorder have already been investigated theoretically, we follow up on these analyses by investigating the changes in the inter-surface forces wrought by the intermediate case, where 
it is assumed that there is a clear separation of relaxation time scales between the dynamics of the ``external" surface charges 
 ($\tau_{\mathrm{s}}$) and the mobile charges  floating in the solvent between the surfaces ($\tau_{\mathrm{ci}}$), {\em i.e.}
\begin{equation}
\label{time_scales}
\tau_{\mathrm{s}}\gg \tau_{\mathrm{ci}}. 
\end{equation}
Thus, the fluctuations in the intervening Coulomb fluid have time enough to relax to their local equilibrium state for each configuration of the surface charge disorder, which shows a much slower relaxation with changes in the surface
separation. The origin of the different relaxation times for the surface and bulk dynamics could be manyfold: the finite mobility and mixing of charged units on the macromolecular surfaces, such as appears in lipid bilayers with embedded charged proteins \cite{lipowsky}, conformational rearrangement of strongly charged DNA chains such as appears in DNA microarrays \cite{science95} and charge regulation of contact surfaces bearing weak acidic groups in aqueous solutions \cite{ParsegianChan}, but should be in any case very {\em system specific}. This separation of time scales  is closely related to the so-called ``adiabatic elimination" of fast degrees of freedom \cite{haken,Risken,Kaneko}, which has been used frequently in the literature 
for systems with widely different time scales \cite{dotsenko1,dotsenko2,landauer,two-T,cool}   (see also Appendix \ref{app:A}). 

In this work we thus investigate the ``interaction fingerprint" of the partially annealed disorder on two apposed planar 
surfaces upon their approach.  The analysis presented above points to the fact that partial annealing of the surface charges  invariably leads to additional attractive interactions between the surfaces and may even result in
a global attractive  instability in the system. The nature of these attractions is quite different, however, if counterions are {\em strongly} or {\em weakly} coupled to charged surfaces  in the sense of Netz \cite{SC_review,netz}. The magnitude of this coupling essentially depends on the valency of counterions, $q$, the magnitude of the surface charge density,
$\sigma e_0$,  and the Bjerrum length $\ell_{\mathrm{B}}=e_0^2/(4\pi \varepsilon\varepsilon_0 k_{\mathrm{B}}T)$ 
(incorporating the medium temperature and dielectric constant), and is measured by the electrostatic
coupling parameter 
\begin{equation}
  \Xi = 2\pi q^3 \ell_{\mathrm{B}}^2 \sigma. 
\end{equation}

For weakly coupled counterions, {\em i.e.} specifically in the mean-field  limit ($\Xi\rightarrow 0$), the 
partially annealed disorder leads to smaller mean-field repulsions due to a renormalized (reduced) value of the surface charge density. The attraction in this case can thus be inferred only from a diminished repulsion with respect to the case of a 
non-disordered surface charge distribution.  For strongly coupled counterions, {\em i.e.}, 
on the SC level  ($\Xi\rightarrow \infty$), we derive explicitly an additional inter-surface attraction 
stemming from the partial annealing of the surface charge distribution.  
Note that the qualitative difference between the mean-field and the SC results
goes back to the strong electrostatic correlation effects that are included in the
SC theory but excluded on the mean-field level. 
The reason for this additional attraction (or reduction of the inter-surface repulsion in the mean-field limit)
compared to  the purely quenched case  is that any rearrangement of the macroion charges, such  as is assumed in partial annealing, inevitably leads to configurations of  lower (free) energy which shows up as an effective attraction. 
This means that  every disordered charged system will be unstable against annealing. In both the mean-field as well as the strong-coupling limit,  the effect of partial annealing is  quantified by a single partial annealing parameter 
 \begin{equation}
\gamma = \frac{ng}{q \sigma},
\end{equation}
which depends on the temperature ratio $n=T/T'$. This is in addition to the disorder coupling parameter 
\begin{equation}
\chi = 2 \pi q^2 \ell_{\mathrm{B}}^2 g, 
\end{equation} 
which measures  the spread of the charge disorder distribution {\em via} the disorder variance $g$. 

The parameter $\gamma$ and the effective surface temperature $T'$ may be estimated experimentally from the mobility ($\Gamma_s$) and the diffusion ($D_s$) coefficients of the surface charges and by applying 
Einstein's relation $k_{\mathrm{B}}T' = D_s/\Gamma_s$ (Appendix \ref{app:elimination}). 
The temperature ratio quantifying the effects of partial annealing can thus be cast into an equivalent form 
\begin{equation}
 n = (k_{\mathrm{B}}T) \frac{\Gamma_s}{ D_s}.
 \end{equation} 
 
 How does one differentiate the effects of quenched and partially annealed disorder? Comparing the results derived in the quenched case  \cite{naji_podgornik,Rudi-Ali} with those obtained in this work, one realizes that in the former case the effects of the disorder are limited to the strong-coupling limit, while in the partially annealed case they persist also in the mean-field limit.
 This is because in the mean-field limit the surface charge density is renormalized in the presence of 
 partially annealed disorder. This effect is always absent in the quenched disorder case (Appendix \ref{app:eff_charge}). 
 
Moreover, we find that in the SC regime and in the limit of low screening $\kappa\rightarrow 0$, a charged system that may 
be stable in the presence of quenched charge disorder could become globally 
unstable and  collapse upon partial annealing of the disorder.  One should also note that for small non-zero
screening and in the stable phase ($\chi<1$), the system adopts a much smaller surface-surface 
distance in the partially annealed case
than in the quenched case as the ratio between the optimal distances in these two cases 
is given by $ (1 +4\gamma/\kappa\mu)^{-1}$.  

Another indicative feature of partially annealed disorder is the
curious property that the surfaces interact electrostatically even 
 if the mean surface charge density, $\sigma$, goes to zero, that is when
 the surfaces become {\em net} electroneutral! It 
 may be  seen  most clearly from the ${\mathrm{Tr}}(\{\hat g\hat v_s\}^2)$ term 
in Eq. (\ref{eq:Trgv^2}), which in the limit $\kappa\rightarrow 0$ leads to 
 a logarithmic attractive contribution in the free energy as $\simeq n \,\pi\, g^2\, \ell_{\mathrm{B}}^2 S \, \ln d$  stemming only
 from the variance, $g$, of the surface charge disorder (note however that when $\sigma>0$ this is 
 a subleading term compared to the additive $1/\kappa$ term 
 considered in Eq. (\ref{eq:SCfree})) \cite{note:threshold}. 
 This effect disappears in the quenched limit ($n\rightarrow 0$)
 unless either the dielectric discontinuities at bounding surfaces or the presence of salt in between the bounding surfaces are taken into account \cite{Rudi-Ali}. 
 %It also vanishes in the mean-field limit (within our Gaussian disorder model). 
 %This is perhaps the most fundamental fingerprint of the disorder. 
 %This fingerprint would allow an experimental differentiation between surfaces with decorated charge distributions. 
 %For net electroneutral surfaces in the case of quenched disorder going from monovalent (weak coupling) to multivalent (strong coupling) counterions would not make any difference. For partially annealed disorder the situation would be as follows: no interaction in the monovalent case, attarction in the multivalent case.  
% added salt and imgae charges change this situation and make it rich
 It is thus safe to say that the effects of partially annealed 
disorder are in general stronger and ubiquitous and may be qualitative in both mean-field and strong-coupling limits.

Nevertheless, though the above-mentioned interaction fingerprints would help in assessing the importance of disorder or even its presence in the charge distribution on macromolecular surfaces, the interactions by themselves would not be enough to make this conclusion with a reasonable degree of certainty.  Unfortunately, more detailed experiments where one would concurrently measure interactions as well as surface charge distributions would still be essential.

Our analysis of the effects of macromolecular charge disorder on their interactions supplements the recently acquired new wisdom of bio-colloidal interactions \cite{SC_review,netz,shkl,levin}, as opposed to its classical formulation \cite{DLVO,Israelachvili}, in quite an illuminating way. Whereas on the DLVO or mean-field level \cite{DLVO,Israelachvili,Andelman} one can formulate the salient features of macromolecular Coulomb interactions with the folk wisdom that opposites attract and likes repel, the strong-coupling paradigm \cite{SC_review,netz,shkl,levin} suggests that 
likes attract too if the system is highly charged. To this we would add, in view of our previous work \cite{naji_podgornik,Rudi-Ali} and the work described above,  that if the surface charge distribution is disordered, the system may become unstable and collapse
due to attractive disorder-induced forces and that even neutral macromolecular surfaces can interact {\em via} 
electrostatic interactions. This is especially true if the disorder distribution is partially annealed as we set to prove in this work.

%%%%%%%%%%%%%%%%%%%%%%%%%%%%%%%%%%%%%%%%%%%%%%%%%%%%%%%%%%%%%
%%%%%%%%%%%%%%%%%%%%%%%%%%%%%%%%%%%%%%%%%%%%%%%%%%%%%%%%%%%%%
\begin{acknowledgements}
R.P. would like to acknowledge the support of Agency for Research and Development of Slovenia
under grants P1-0055(C), Z1-7171  and L2-7080. This study was supported by the Intramural Research Program of the NIH, National Institute of Child Health and Human Development. This research was supported in part by the National Science Foundation under Grant No. PHY05-51164.
\end{acknowledgements}

%%%%%%%%%%%%%%%%%%%%%%%%%%%%%%%%%%%%%%%%%%%%%%%%%%%%%%%%%%%%%
%%%%%%%%%%%%%%%%%%%%%%%%%%%%%%%%%%%%%%%%%%%%%%%%%%%%%%%%%%%%%
\appendix
\section{Dynamics of disorder}
\label{app:A}

\subsection{Adiabatic elimination of fast variables}
\label{app:elimination}

The effective partition function for a partially annealed system may be obtained {\em via} an adiabatic
elimination of fast variables discussed widely in the literature \cite{haken,Risken,Kaneko}. Here we shall briefly discuss the 
extension of this method for a charged system with a slow disordered component. As described in the text we 
shall focus on a non-equilibrium system of mobile fast-relaxing counterions (located at positions ${\boldsymbol \xi}_i(t)$ for
$i=1,\ldots,N$, in a solvent of fixed 
temperature $T$) and a disordered ensemble of 
macroion surface charges (located at positions ${\mathbf x}_i$ with $i=1,\ldots,N_{\mathrm{s}}$ and bearing
charges $q_{\mathrm{s}}e_0$) which undergo 
a much slower dynamics. We shall be interested only in the stationary-state properties of the system. 
We adopt the Langevin equation for the dynamics of counterions
and surface charges as
\begin{eqnarray}
  \dot {\boldsymbol \xi}_i &=& -\Gamma_{\mathrm{ci}} \frac{\partial H}{\partial {\boldsymbol \xi}_i} + {\boldsymbol \theta}_{\mathrm{ci}}, 
  \label{eq:Langevin_xi_i}
  \\
  \dot {\mathbf x}_i &=& -\Gamma_{\mathrm{s}} \frac{\partial H}{\partial {\mathbf x}_i} + {\boldsymbol \theta}_{\mathrm{s}}, 
  \label{eq:Langevin_x_i}
\end{eqnarray}
respectively, where the Hamiltonian $H$ is written  in terms of Coulomb pair interactions
\begin{equation}
  H = \frac{1}{2}\sum_{i,j} v({\mathbf x}_i, {\mathbf x}_j) + \sum_{i,j} v({\mathbf x}_i, {\boldsymbol \xi}_j) +
  \frac{1}{2}  \sum_{i,j} v({\boldsymbol \xi}_i, {\boldsymbol \xi}_j), 
\end{equation} 
and ${\boldsymbol \theta}_{\mathrm{ci}}$ and ${\boldsymbol \theta}_{\mathrm{s}}$ are white noise terms 
with zero average and the two-point correlation function $\overline{ {\boldsymbol \theta}_{\mathrm{ci}}(t)\cdot{\boldsymbol \theta}_{\mathrm{ci}}(t')} = 
6D_{\mathrm{ci}} \, \delta(t-t')$ and $\overline{ {\boldsymbol \theta}_{\mathrm{s}}(t)\cdot{\boldsymbol \theta}_{\mathrm{s}}(t')} = 
6D_{\mathrm{s}} \, \delta(t-t')$. We impose Einstein's relations between the diffusion constants and the mobilities of the
particle species, {\em i.e.},  $D_{\mathrm{ci}} =  k_{\mathrm{B}} T \Gamma_{\mathrm{ci}} $
and $D_{\mathrm{s}} =  k_{\mathrm{B}} T' \Gamma_{\mathrm{s}}$ that guarantee a  
stationary state may be achieved at long times. (We have neglected hydrodynamic interactions in Eqs. (\ref{eq:Langevin_xi_i})
and (\ref{eq:Langevin_xi_i}) for simplicity as they do not affect the stationary-state results.) 

The central assumption for the time scales of the 
counterions ($1/\Gamma_{\mathrm{ci}}$) and the surface charges ($1/\Gamma_{\mathrm{s}}$) translates into 
\begin{equation}
\Gamma_{\mathrm{s}}/\Gamma_{\mathrm{ci}} \ll 1.
\label{eq:eq_GammaRatio} 
\end{equation}
Our goal is to eliminate the the fast (counterionic) variables that change at short 
time scales and obtain an {\em effective} dynamical equation for the slow (surface charge) variables assuming
that the effective temperature of the latter, $T'$, 
 may in general be different from that of 
the former, $T$ \cite{dotsenko1,dotsenko2,landauer,two-T,cool}.
This may be done readily given the condition (\ref{eq:eq_GammaRatio}) above and
by developing a perturbative expansion (much in the 
spirit of the Born-Oppenheimer approximation) \cite{haken,Risken,Kaneko}
in terms of the small parameter 
$\Gamma_{\mathrm{s}}/\Gamma_{\mathrm{ci}}$. 
The zeroth-order results are quite intuitive: one ends up with a Langevin equation for
${\mathbf x}_i $ similar to Eq. (\ref{eq:Langevin_x_i}) where the microscopic Hamiltonian $H$ is replaced 
with an effective one, which is obtained by ``pre-averaging" over the counterionic degrees of freedom; namely, 
\begin{equation}
  \dot {\mathbf x}_i = -\Gamma_{\mathrm{s}} \frac{\partial H_{\mathrm{eff}}}{\partial {\mathbf x}_i} + {\boldsymbol \theta}_{\mathrm{s}}, 
  \label{eq:Langevin_eff}
\end{equation}
where $H_{\mathrm{eff}} =  -k_{\mathrm{B}}T\ln Z_{\mathrm{ci}}$ and
$Z_{\mathrm{ci}} = \frac{1}{N!}\int \prod_i{\mathrm{d}}{\boldsymbol \xi}_i\, \exp(-\beta H)$. 
This effective Hamiltonian is of course nothing but the equilibrium free energy of counterions 
in the presence of a {\em fixed} realization of surface charges; its derivative acts as
a driving force pushing the surface charge dynamics  to reach a stationary state at long times
 with the Boltzmann-type probability weight $\exp(-\beta' H_{\mathrm{eff}})$ as dictated
by Eq. (\ref{eq:Langevin_eff}). Hence, the system--although far from equilibrium as a whole--may 
still be described in terms of an ``effective" partition  function, $Z$, once a stationary state
is achieved, {\em i.e.} 
\begin{equation}
  Z =  \frac{1}{N_{\mathrm{s}}!}\int \bigg[\prod_i{\mathrm{d}}{\mathbf x}_i\bigg]\,\, e^{-\beta' H_{\mathrm{eff}}} = 
     \frac{1}{N_{\mathrm{s}}!} \int \bigg[\prod_i{\mathrm{d}}{\mathbf x}_i\bigg]\,\, \left(Z_{\mathrm{ci}}\right)^n, 
      \label{eq:Z_eff}
\end{equation}
where $n=\beta'/\beta$. 

\subsection{Dynamical density functional theory}

It is possible to derive the $n$-replica partition function (\ref{part_fn}) directly from 
the expression (\ref{eq:Z_eff}) and by using a Legendre transformation to grand-canonical 
ensemble described by the partition functions 
${\mathcal Z}=\sum_{N_{\mathrm{s}}} \lambda_{\mathrm{s}}^{N_{\mathrm{s}}}\, Z$ and  
${\mathcal Z}_{\mathrm{ci}}=\sum_N \lambda^N\, Z_{\mathrm{ci}}$, where
$ \lambda_{\mathrm{s}}$ and $ \lambda$ denote the fugacities of the surface charges and
counterions, respectively. % cite future work
We shall however take a detour over the so-called dynamical density functional theory that
first transforms the effective Langevin equation (\ref{eq:Langevin_eff}) into a dynamical equation
for the surface charge density field $\rho({\mathbf r}, t)$. This may be done straightforwardly following the approach proposed for classical fluids in Ref. \cite{DDFT}. One thus obtains
\begin{equation}
 \frac{\partial}{\partial t} \rho({\mathbf r}, t) = - \nabla\cdot {\mathbf J}_{\rho}({\mathbf r}, t) + {\mathbf \eta} ({\mathbf r}, t), 
 \label{eq:continuity}
\end{equation}
assuming a conserved density field \cite{Halperin,ma}, 
where the current density reads
\begin{equation}
 {\mathbf J}_{\rho}({\mathbf r}, t) = -D_{\mathrm{s}}\, \nabla {\rho}({\mathbf r}, t) - (q_{\mathrm{s}}e_0\Gamma_{\mathrm{s}}) \, 
 	 {\rho}({\mathbf r}, t)\, \nabla \langle \psi ({\mathbf r}, t) \rangle_T.
	  \label{eq:current}		
\end{equation}
The first term on the right hand side represents the current due to the diffusion of surface charges
and the second term represents the contribution from the mean effective electrostatic force acting on them;
here  $ \langle \psi ({\mathbf r}, t) \rangle_T =  -\delta \ln {\mathcal Z}_{\mathrm{ci}}[\rho]/\beta \delta \rho({\mathbf r})$ is 
the electrostatic potential at the {\em macroion surface} averaged over the counterionic degrees of freedom. 
The last term is a Gaussian white noise with 
$\overline{\eta({\mathbf r}, t)}=0$ and $\overline{ \eta({\mathbf r}, t) \,\eta({\mathbf r}', t')} 
= 2 \nabla\cdot\nabla' [D_{\mathrm{s}}\, q_{\mathrm{s}} e_0 \, {\rho}({\mathbf r}, t)\,\delta({\mathbf r}-{\mathbf r}') ]\, \delta(t-t')$.
% Note that the emergence of a multiplicative noise ({\em i.e.} with noise variance being dependent
% on the local density) is a natural consequence of the fact that the
% local density does not fluctuate in empty regions
Combining Eqs. (\ref{eq:continuity}) and (\ref{eq:current}), one arrives at 
\begin{eqnarray}
 \frac{\partial}{\partial t} \rho({\mathbf r}, t) &=&  \nabla\cdot \bigg( D_{\mathrm{s}}\, \nabla {\rho}
 		+   (q_{\mathrm{s}} e_0 \Gamma_{\mathrm{s}})\, \rho\, \nabla \langle \psi \rangle_T\bigg) + \eta({\mathbf r}, t)\\
		&=&  \nabla\cdot \bigg[ ( q_{\mathrm{s}} e_0  \Gamma_{\mathrm{s}}) \, \rho({\mathbf r}, t)\, \nabla
 										\frac{\delta {\mathcal W}[\rho]}{\delta \rho ({\mathbf r}, t)} \bigg]
						+ \eta({\mathbf r}, t),
		\label{eq:langevin_W}
\end{eqnarray}
where the effective ``potential", ${\mathcal W}[\rho]$, follows as
\begin{equation}
 {\mathcal W}[\rho] = \frac{1}{q_{\mathrm{s}} e_0 \beta'} \int {\mathrm{d}} {\mathbf r}\, \bigg[ \rho\, \ln \bigg(\frac{\rho}{\rho_0}\bigg) - \rho  \bigg]
 				- \frac{1}{\beta} \, \ln {\mathcal Z}_{\mathrm{ci}}[\rho]
		\label{eq:Weight}		
\end{equation}
with $\rho_0$ being a constant. 
Using a standard mapping of Eq. (\ref{eq:langevin_W}) to an equivalent Fokker-Planck equation, 
one can easily show that the stationary-state fluctuations
of $\rho({\mathbf r}, t)$ follow a Boltzmann-type weight $\exp(-\beta'  {\mathcal W})$ and hence, 
the effective ``partition function"
\begin{equation}
 {\mathcal Z} =  \int {\mathcal D}\rho \,\,\,\, e^{ -  \frac{1}{q_{\mathrm{s}} e_0}
 				\int {\mathrm{d}} {\mathbf r}\, [\rho \, \ln (\rho/\rho_0) - \rho]}
			 			\,\bigg(  {\mathcal Z}_{\mathrm{ci}}[\rho({\mathbf r})]  \bigg)^n.
\label{eq:DFT_Z_Zci_n}
\end{equation}

In order to proceed with an analytical investigation, we assume that the surface density fluctuations 
around $\rho_0$ are small and expand the entropic term (first term in Eq. (\ref{eq:Weight})) up to the second order, which
gives $\rho \, \ln (\rho/\rho_0) - \rho \simeq (\rho - \rho_0)^2/2\rho_0$. 
One can generalize this result by assuming that different realizations of the surface charge disorder 
follow a Gaussian probability distribution as 
\begin{equation}
   {\mathcal P}[\rho] = C\, \exp\bigg(\!\! - \frac{1}{2}  \int {\mathrm{d}} {\mathbf r}\,  g^{-1}({\mathbf r})\,
   								[\rho({\mathbf r}) - \rho_0({\mathbf r}) ]^2 \bigg),
\end{equation}
where %$C = [2\pi g({\mathbf r})]^{-1/2}$ 
$C$ is a normalization factor 
and $g({\mathbf r})$ is the disorder variance. The quadratic {\em ansatz} is completely general since it corresponds to an expansion in small deviations from  $\rho_0({\mathbf r})$ where $g({\mathbf r})$ can be interpreted as a disorder ``compressibility". One thus recovers Eqs. (\ref{eq:eff_pot}) and (\ref{eq:Z_Zci_n}) in the text.

\subsection{Non-conserved disorder dynamics}

The dynamical model discussed above is based on a conserved model for
the dynamics of the disorder, where the current density follows 
from the gradient of an effective chemical potential, {\em i.e.}, 
${\mathbf J}_{\rho} = -( q_{\mathrm{s}} e_0\Gamma_{\mathrm{s}}) \, \rho\, \nabla \mu_{\mathrm{eff}}[\rho]$
with 
\begin{equation}
\mu_{\mathrm{eff}}[\rho] \equiv \frac{\delta {\mathcal W}[\rho]}{\delta \rho} = \frac{1}{q_{\mathrm{s}} e_0\beta'} \ln \bigg(\frac{\rho}{\rho_0}\bigg) 
			+ \langle \psi ({\mathbf r}, t) \rangle_T
\end{equation}
as seen from Eq. (\ref{eq:langevin_W}). 
Alternatively, one could construct other dynamical
models that include a non-conserved density field for the surface charge disorder and still
give rise to the same effective partition function in the stationary-state limit \cite{Halperin,ma}. One may thus propose
a phenomenological equation of the form 
\begin{equation}
 \frac{\partial}{\partial t} \rho({\mathbf r}, t) = 
 	- \tilde \Gamma_{\mathrm{s}} \big( \mu_{\mathrm{eff}}[\rho] -  \mu_0\big) + {\mathbf \eta} ({\mathbf r}, t), 
\end{equation}
where $\tilde \Gamma_{\mathrm{s}}$ is an effective mobility, $ \mu_0$ is a constant (which can
be set to zero) and $\eta$ is a Gaussian white noise with 
$\overline{\eta({\mathbf r}, t)}=0$ and $\overline{ \eta({\mathbf r}, t) \,\eta({\mathbf r}', t')} 
= 2 \tilde D_{\mathrm{s}} \, \delta({\mathbf r}-{\mathbf r}')\, \delta(t-t')$.
Assuming $\tilde D_{\mathrm{s}} = k_{\mathrm{B}}T' \tilde \Gamma_{\mathrm{s}}$ one can show that
the partition function (\ref{eq:DFT_Z_Zci_n}) is reproduced as the stationary-state solution.
This model may be more appropriate to study dynamics of disorder in macromolecules with ionizable surface groups.

\section{Partially annealed free energy}
\label{app:B}

The ``free energy" of a partially annealed disordered system follows from Eq. (\ref{eq:Z_Zci_n}) as
\begin{equation}
\label{eq:part_ann_freen}
{\mathcal F} = - k_{\mathrm{B}}T' \ln  {\mathcal Z}
	= - k_{\mathrm{B}}T' \ln \, \langle   \!\! \langle   {\mathcal Z}_{\mathrm{ci}} ^n  \rangle   \!\! \rangle.
\end{equation}
Obviously, the  purely annealed disorder limit with true thermal equilibrium, $T=T'$, is recovered
when $n=1$, 
\begin{equation}
\label{eq:annealed_freen}
{\mathcal F}_{\mathrm{annealed}}
	= - k_{\mathrm{B}}T\ln \, \langle  \! \! \langle   {\mathcal Z}_{\mathrm{ci}}   \rangle   \!\! \rangle,
\end{equation}
while the purely quenched disorder limit follows by taking the limit $n\rightarrow 0$
(or $\beta'  \rightarrow 0$).
%, signifying that the surface and the bulk are completely decoupled from one another.
One can use  $\langle   \!\! \langle   {\mathcal Z}_{\mathrm{ci}} ^n  \rangle   \!\! \rangle = e^{-\beta' {\mathcal F} }$
and expand both sides of this relation to obtain the standard quenched free energy expression \cite{dotsenko1,dotsenko2}
\begin{equation}
\label{eq:quenched_freen}
{\mathcal F}_{\mathrm{quenched}}
	= - k_{\mathrm{B}}T \langle \!  \! \langle  \ln {\mathcal Z}_{\mathrm{ci}}  \rangle \!  \! \rangle.
\end{equation}
All other values of $n$ represent a partially annealed situation. 
The quenched limit translates into an infinite temperature for the disorder  \cite{dotsenko1,dotsenko2,cool}
relative to counterions, reflecting the fact that the statistics of quenched disorder is unaffected 
by counterions. 
It also reflects the zero mobility of surface charges when the dynamical interpretation {\em via} Einstein's relation
 is used (assuming a finite surface diffusion coefficient). 
%Let us not forget here that the surface temperature $T'$ reflects the surface dynamics {\em via} the Einstein relation and is proportional to the ratio of the surface diffusion constant to the mobility of the surface charges.

\section{Effective surface charge density}
\label{app:eff_charge}

The effective surface charge density $\sigma_{\mathrm{eff}}$ in the two-plate system is defined via 
\begin{equation}
  \sigma_{\mathrm{eff}} = -\frac{1}{2S e_0}\int{\mathrm{d}}{\mathbf r} \,\,[[\rho({\mathbf r})]],
  \label{eq:app_sigma_eff}
 \end{equation}
 where $[[\cdots]] = \frac{1}{\mathcal Z} \int {\mathcal D}\rho\, {\mathcal P}[\rho]\,\big(  {\mathcal Z}_{\mathrm{ci}}[\rho]  \big)^n \, (\cdots)$ is the full ensemble average with respect to the partition ${\mathcal Z}$, Eq. (\ref{eq:Z_Zci_n}). It may
 be written in terms of $ \langle \!\! \langle  \cdots \rangle   \!\! \rangle$ averages defined after Eq. (\ref{eq:Z_Zci_n}) as 
 $[[\cdots]] = \langle \!\! \langle  \big(  {\mathcal Z}_{\mathrm{ci}}[\rho]  \big)^n\, (\cdots) \rangle   \!\! \rangle/\langle \!\! \langle  \big(  {\mathcal Z}_{\mathrm{ci}}[\rho]  \big)^n \rangle   \!\! \rangle$. 
One can easily show from Eqs. (\ref{eq:Z_Zci_n}) and (\ref{part_fn}) that 
\begin{equation}
  [[\rho({\mathbf r})]] = \rho_0({\mathbf r}) + g({\mathbf r})\frac{\delta\ln {\mathcal Z}}{\delta \rho_0({\mathbf r})}
				= \rho_0({\mathbf r}) - {\mathrm{i}}\, \beta\, g({\mathbf r})\sum_a\, [[\phi_a({\mathbf r})]]. 
\label{eq:sigma_eff_id}
\end{equation}
Obviously, the effective or renormalized surface charge is shifted from the bare value, $\rho_0({\mathbf r})$, by an
amount proportional to the mean surface potential $ [[\phi_a({\mathbf r})]]$. 
Thus the charge renormalization in the present context is rendered {\em via} a linear surface screening
mechanism (as may be seen also from the quadratic disorder terms $g({\mathbf r}) \phi_a({\mathbf r}) \phi_b({\mathbf r})$ 
in the effective Hamiltonian (\ref{ham})).   This can be traced back to the Gaussian approximation for the 
disorder distribution, Eq. (\ref{eq:gaussian}).

It follows immediately from Eq. (\ref{eq:sigma_eff_id}) that no charge renormalization occurs in the quenched limit ($n\rightarrow 0$)
and it may be possible only for a partially annealed disorder ($n>0$). 
Assuming the replica symmetry {\em ansatz}, one can recover the PB effective charge, Eqs. (\ref{eq:PB_rho_eff}) and
(\ref{renormedsigma}), in the mean-field limit. In the SC limit, the effective charge and the fugacity, $\lambda$, are
determined simultaneously from Eqs. (\ref{eq:app_sigma_eff}) and (\ref{eq:sigma_eff_id}) above combined with 
Eq. (\ref{eq:N_lambda}) or (\ref{eq:lambda}) in the text and the electroneutrality 
condition $N q = 2 \sigma_{\mathrm{eff}} S$, {\em i.e.}
\begin{eqnarray}
   \lambda &=& \frac{2 \sigma_{\mathrm{eff}} S}{q\int {\mathrm{d}}{\mathbf R} \, \Omega({\mathbf R})\, e^{-\beta u({\mathbf R})}}, \\
    \sigma_{\mathrm{eff}} & =& -\frac{1}{2S e_0}\int{\mathrm{d}}{\mathbf r}\, \bigg[ \rho_0({\mathbf r}) + g({\mathbf r})\frac{\delta\ln {\mathcal Z}}{\delta \rho_0({\mathbf r})}\bigg]. 
\end{eqnarray}
By using the SC results for the single-particle interaction, $u$, and the partition function, ${\mathcal Z}$, from
Sections  \ref{subsec:virial} and \ref{subsec:small_n}, and expanding $ \sigma_{\mathrm{eff}} $ and 
$\lambda$ in powers of $n$ as 
\begin{eqnarray}
  \sigma_{\mathrm{eff}} &=& \sigma^{(0)} + n\, \sigma^{(1)}+\cdots,\\
  \lambda &=& \lambda^{(0)} + n\, \lambda^{(1)}+\cdots, 
\end{eqnarray}
we find that $ \sigma^{(0)} = \sigma$, $ \lambda^{(0)} = 2 \sigma S/q\int {\mathrm{d}}{\mathbf R} \, \Omega({\mathbf R})\, e^{-\beta u_0({\mathbf R})}$, and
\begin{eqnarray}
    \frac{\sigma^{(1)}}{\sigma} &=& 
      \frac{\beta}{2\sigma Se_0}\int{\mathrm{d}}{\mathbf r}\, g({\mathbf r})\, \bigg[\langle \rho_0|\hat v_s|{\mathbf r}\rangle + 
    	 q e_0  \,\lambda^{(0)} \int {\mathrm{d}}{\mathbf R} \, \Omega({\mathbf R})\, e^{-\beta u_0({\mathbf R})}
		\langle {\mathbf r}|\hat v_s|{\mathbf R} \rangle\bigg], \\
        \frac{\lambda^{(1)}}{\lambda^{(0)}} & = &\frac{\sigma^{(1)}}{\sigma} 
        			- \beta \frac{\int {\mathrm{d}}{\mathbf R} \, \Omega({\mathbf R})\, u_1({\mathbf R})\, e^{-\beta u_0({\mathbf R})} }
    					  {\int {\mathrm{d}}{\mathbf R} \, \Omega({\mathbf R})\, e^{-\beta u_0({\mathbf R})} }. 
\end{eqnarray}
The $\lambda^{(1)}$ term and other higher-order terms lead to higher-order contributions 
in the effective charge and thus will not be relevant here. 
By straightforward calculations and using the results in the text, one can show that 
\begin{equation}
    \frac{\sigma^{(1)}}{\sigma} = g\ell_{\mathrm{B}}\,\left(\frac{2\pi}{\kappa}\right)\bigg[-(1+e^{-\kappa d}) + 
        \frac{\int {\mathrm{d}}{\mathbf R} \, \Omega({\mathbf R})\, e^{-\beta u_0({\mathbf R})} 
        		\big(e^{-\kappa|a-R_z|}+e^{-\kappa|a+R_z|}\big)}
        				{\int {\mathrm{d}}{\mathbf R} \, \Omega({\mathbf R})\, e^{-\beta u_0({\mathbf R})}}\bigg]. 
\end{equation}
In the limit $\kappa\rightarrow 0$, the first two terms in the bracket on the 
right hand side are cancelled by similar contributions from the integrals in the last term and one can show that 
\begin{equation}
   \lim_{\kappa\rightarrow 0} \frac{\sigma^{(1)}}{\sigma} =0. 
\end{equation}
Therefore, in the SC limit, the surface charge density is not renormalized on the leading order in $n$ when the
low screening limit is taken. Thus, we can impose the electroneutrality condition using the bare surface charge density 
as $ 2 \sigma S = q N$.

%%%%%%%%%%%%%%%%%%%%%%%%%%%%%%%


\begin{thebibliography}{99}
\bibitem{note:threshold}
Subleading terms of order ${\mathcal O}(n\, \kappa^0)$ can in general result in a small shift in the threshold value
of $\chi$. For instance, the ${\mathrm{Tr}}(\{\hat g\hat v_s\}^2)$ term, Eq. (\ref{eq:Trgv^2}), leads to a small reduction 
of the order ${\mathcal O}(n)$ as  $\chi_c \simeq 1- \gamma/4$.
\end{thebibliography}
\end{document}